\documentclass[12pt,onecolumn]{IEEEtran}

\ifCLASSINFOpdf
   \usepackage[pdftex]{graphicx}
\else
\fi
\usepackage[cmex10]{amsmath}
\usepackage{amssymb}   % \triangleq vs. gibi seyler bunda
\usepackage{amsxtra}
\usepackage{amscd}
\usepackage{amsthm}

\usepackage{graphicx}   % sekilller icin

\usepackage{array}  % tablo icin satir acma

\usepackage{multirow}  % tablolar icin coklu satir kullanmada

\usepackage{flushend} % son sayfa cift kolon yapiyor

\usepackage{cite}

\begin{document}
%
% paper title
% can use linebreaks \\ within to get better formatting as desired
\title{{\huge Multiple-Input Multiple-Output OFDM with Index Modulation }}

%\title{{\LARGE Multiple-Input Multiple-Output OFDM with Index Modulation: \\ An Alternative Multicarrier Transmission Scheme for 5G Networks }}

%
%
% author names and IEEE memberships
% note positions of commas and nonbreaking spaces ( ~ ) LaTeX will not break
% a structure at a ~ so this keeps an author's name from being broken across
% two lines.
% use \thanks{} to gain access to the first footnote area
% a separate \thanks must be used for each paragraph as LaTeX2e's \thanks
% was not built to handle multiple paragraphs
%

\author{Ertu\u{g}rul~Ba\c{s}ar,~\IEEEmembership{Member,~IEEE} 
	%\vspace*{-0.25cm}
	\thanks{E. Ba\c{s}ar is with Istanbul Technical University, Faculty of Electrical and Electronics Engineering, 34469, Istanbul, Turkey. e-mail: basarer@itu.edu.tr. This work was supported by the Scientific Research Projects Foundation, Istanbul Technical University.}
	\thanks{\copyright 2015 IEEE. Personal use of this material is permitted. Permission from IEEE must be obtained for all other uses, in any current or future media, including reprinting/republishing this material for advertising or promotional purposes, creating new collective works, for resale or redistribution to servers or lists, or reuse of any copyrighted component of this work in other works.}
	\thanks{Digital Object Identifier 10.1109/LSP.2015.2475361}	
	}

% note the % following the last \IEEEmembership and also \thanks -
% these prevent an unwanted space from occurring between the last author name
% and the end of the author line. i.e., if you had this:
%
% \author{....lastname \thanks{...} \thanks{...} }
%                     ^------------^------------^----Do not want these spaces!
%
% a space would be appended to the last name and could cause every name on that
% line to be shifted left slightly. This is one of those "LaTeX things". For
% instance, "\textbf{A} \textbf{B}" will typeset as "A B" not "AB". To get
% "AB" then you have to do: "\textbf{A}\textbf{B}"
% \thanks is no different in this regard, so shield the last } of each \thanks
% that ends a line with a % and do not let a space in before the next \thanks.
% Spaces after \IEEEmembership other than the last one are OK (and needed) as
% you are supposed to have spaces between the names. For what it is worth,
% this is a minor point as most people would not even notice if the said evil
% space somehow managed to creep in.

% The paper headers
\markboth{IEEE Signal Processing Letters}{Ba\c{s}ar:  Multiple-Input Multiple-Output OFDM with Index Modulation}

% The only time the second header will appear is for the odd numbered pages
% after the title page when using the twoside option.
%
% *** Note that you probably will NOT want to include the author's ***
% *** name in the headers of peer review papers.                   ***
% You can use \ifCLASSOPTIONpeerreview for conditional compilation here if
% you desire.

% If you want to put a publisher's ID mark on the page you can do it like
% this:
%\IEEEpubid{0000--0000/00\$00.00~\copyright~2007 IEEE}
% Remember, if you use this you must call \IEEEpubidadjcol in the second
% column for its text to clear the IEEEpubid mark.

% use for special paper notices
%\IEEEspecialpapernotice{(Invited Paper)}

% make the title area

\maketitle

\begin{abstract}
%\boldmath
Orthogonal frequency division multiplexing with index modulation (OFDM-IM) is a novel multicarrier transmission technique which has been proposed as an alternative to classical OFDM. The main idea of OFDM-IM is the use of the indices of the active subcarriers in an OFDM system as an additional source of information. In this work, we propose multiple-input multiple-output OFDM-IM (MIMO-OFDM-IM) scheme by combining OFDM-IM and MIMO transmission techniques. The low complexity transceiver structure of the MIMO-OFDM-IM scheme is developed and it is shown via computer simulations that the proposed MIMO-OFDM-IM scheme achieves significantly better error performance than classical MIMO-OFDM for several different system configurations.

\end{abstract}
% IEEEtran.cls defaults to using nonbold math in the Abstract.
% This preserves the distinction between vectors and scalars. However,
% if the journal you are submitting to favors bold math in the abstract,
% then you can use LaTeX's standard command \boldmath at the very start
% of the abstract to achieve this. Many IEEE journals frown on math
% in the abstract anyway.
\vspace*{-0.1cm}
% Note that keywords are not normally used for peerreview papers.

% For peer review papers, you can put extra information on the cover
% page as needed:
% \ifCLASSOPTIONpeerreview
% \begin{center} \bfseries EDICS Category: 3-BBND \end{center}
% \fi
%
% For peerreview papers, this IEEEtran command inserts a page break and
% creates the second title. It will be ignored for other modes.
\IEEEpeerreviewmaketitle

\renewcommand{\thefootnote}{\fnsymbol{footnote}}

\vspace*{-0.17cm}
\section{Introduction}
% The very first letter is a 2 line initial drop letter followed
% by the rest of the first word in caps.
%
% form to use if the first word consists of a single letter:
% \IEEEPARstart{A}{demo} file is ....
%
% form to use if you need the single drop letter followed by
% normal text (unknown if ever used by IEEE):
% \IEEEPARstart{A}{}demo file is ....
%
% Some journals put the first two words in caps:
% \IEEEPARstart{T}{his demo} file is ....
%
% Here we have the typical use of a "T" for an initial drop letter
% and "HIS" in caps to complete the first word.
\IEEEPARstart{O}{rthogonal} frequency division multiplexing (OFDM) has become the most popular multicarrier signaling format for high-speed wireless communications and has been included in many standards such as Long Term Evolution (LTE), IEEE 802.11 wireless local area network (WLAN) and digital video broadcasting (DVB). Due to its efficient implementation and robustness to the frequency selectivity, OFDM and its combination with multiple-input multiple-output (MIMO) systems unsurprisingly  appears as a strong alternative for 5G networks \cite{5G}.

OFDM with index modulation (OFDM-IM) is a recently proposed novel scheme which transmits information not only by $M$-ary constellation symbols, but also by the indices of the active subcarriers which are activated according to the incoming information bits \cite{OFDM_IM}. Subcarrier index modulation techniques for OFDM \cite{OFDM_IM,SIM_OFDM,ESIM_OFDM} have attracted considerable attention from researchers and have been investigated in some recent studies due to interesting trade-offs they offer in error performance and spectral efficiency compared to classical OFDM systems \cite{MCIK_OFDM,OFDM_GIM,Opt_OFDM_IM,GSScM,CI_OFDM_IM,OFDM_ISIM,IM_OFDM_V2X}. The bit error probability of OFDM-IM is analytically derived in \cite{MCIK_OFDM}. The spectral efficiency of OFDM-IM is improved by selecting the active subcarriers in a more flexible way in \cite{OFDM_GIM}, where index modulation is applied for both in-phase and quadrature components of the subcarriers. In \cite{Opt_OFDM_IM} and \cite{GSScM}, the authors deal with the problem of selecting the optimal number of active subcarriers in OFDM-IM. More recently, OFDM-IM is combined with coordinate interleaving to achieve additional diversity gains in \cite{CI_OFDM_IM}. However, the combination of OFDM-IM and MIMO transmission techniques remains an open and interesting research problem.

In this study, we propose MIMO-OFDM with index modulation (MIMO-OFDM-IM) as an efficient alternative multicarrier transmission scheme for 5G networks by combining MIMO and OFDM-IM transmission techniques. In the proposed scheme, each transmit antenna transmits its own OFDM-IM frame as in Vertical Bell Labs layered space-time (V-BLAST) scheme \cite{MIMO_OFDM}, and at the receiver side, these OFDM-IM frames are separated and demodulated using a novel and low complexity minimum mean square error (MMSE) detection and log-likelihood ratio (LLR) calculation based detector. It is shown via computer simulations that the MIMO-OFDM-IM scheme achieves significantly better bit error rate (BER) performance than classical V-BLAST type MIMO-OFDM for several MIMO configurations. 

The rest of the letter is organized as follows. In Section II, the system model of MIMO-OFDM-IM is presented. In Section III, receiver structure of the MIMO-OFDM-IM scheme is given. Simulation results are provided in Section IV. Finally, Section V concludes the letter.\footnote{{\em Notation}: Bold, lowercase and capital letters are used for column vectors and matrices, respectively. $(\mathbf{A})_t$ and $(\mathbf{A})_{t,t}$ denote the $t$th column and the $t$th main diagonal element of $\mathbf{A}$, respectively. ${{\left( \cdot \right)}^{T}}$ and ${{\left( \cdot \right)}^{H}}$ denote transposition and Hermitian transposition, respectively. $\mathbf{I}_{N}$ is the identity matrix with dimensions $N\times N$ and $\mbox{diag}\left(\cdot \right) $ denotes a diagonal matrix. $\left\| \cdot\right\| $ stands for the Euclidean  norm. The probability of an event is denoted by ${{P}}\left( \cdot \right)$ and $E\left\lbrace \cdot \right\rbrace $ stands for expectation. $X\sim \mathcal{CN}\left( 0,\sigma _{X}^{2} \right)$ represents the distribution of a circularly symmetric complex Gaussian r.v. $X$ with variance $\sigma _{X}^{2}$. $C\left(N,K \right) $ stands for the binomial coefficient and $\lfloor\cdot\rfloor$ is the floor function. $\mathcal{S}$ denotes $M$-ary signal constellation. $ \mathbb{C} $ denotes the ring of complex numbers.}  

%\vspace*{-0.15cm}
\section{System Model of MIMO-OFDM-IM}
\begin{figure*}[t]
\begin{center}
{\includegraphics[scale=1.0]{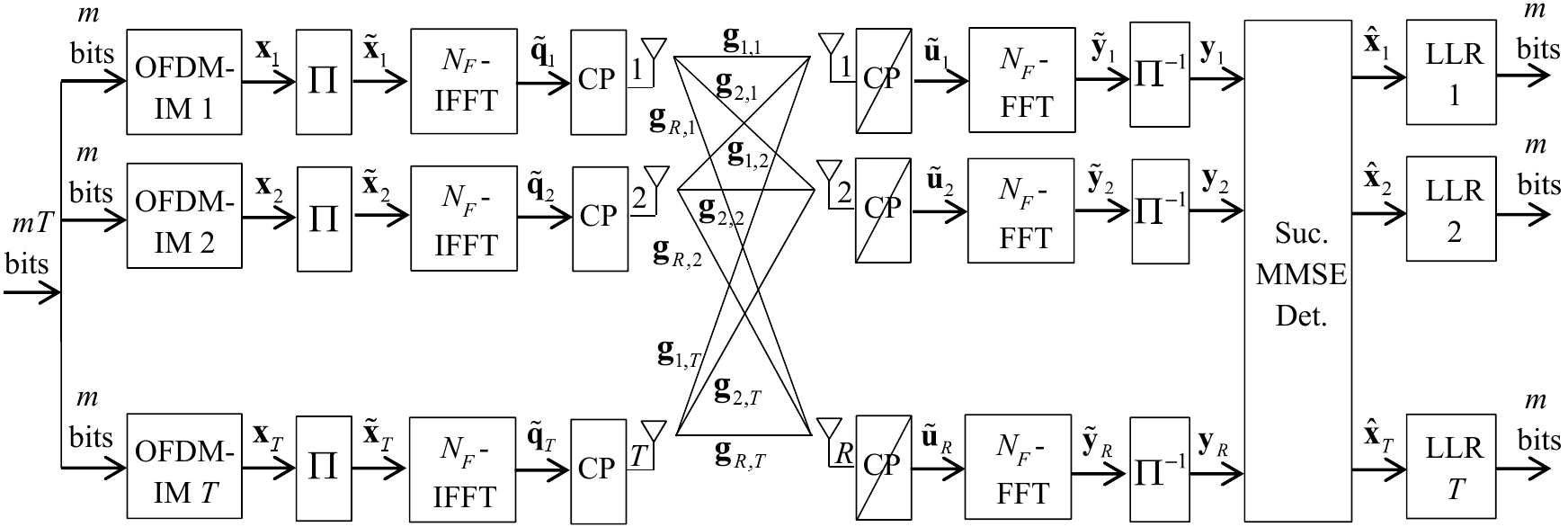}}
%\vspace*{-0.2cm}
\caption{Transceiver Structure of the MIMO-OFDM-IM Scheme for a $T\times R$ MIMO System}
%\vspace*{-0.7cm}
\end{center}
\end{figure*}

The block diagram of the MIMO-OFDM-IM transceiver is shown in Fig. 1. We consider a MIMO system employing $T$ transmit and $R$ receive antennas. As seen from Fig. 1, for the transmission of each frame, a total of $mT$ information bits enter the  MIMO-OFDM-IM transmitter. These $mT$ bits are first split into $T$ groups and the corresponding $m$ bits are processed in each branch of the transmitter by the OFDM index modulators. The incoming $m$ information bits are used to form the $N_F \times 1$ OFDM-IM block $\mathbf{x}_t=\begin{bmatrix}
x_t(1) & x_t(2) &\cdots &x_t(N_F)
\end{bmatrix}^T,t=1,2,\ldots,T$ in each branch of the transmitter, where $N_F$ is the size of the fast Fourier transform (FFT) and $x_t(n_f) \in \left\lbrace 0,\mathcal{S} \right\rbrace,n_f=1,2,\ldots,N_F $. According to the OFDM-IM principle \cite{OFDM_IM}, %which is carried out simultaneously in each branch of the transmitter, 
these $m$ bits are split into $G$ groups each containing $p=p_1 + p_2$ bits, which are used to form OFDM-IM subblocks $\mathbf{x}_t^g=\begin{bmatrix}
x_t^g(1) & x_t^g(2) &\cdots &x_t^g(N)
\end{bmatrix}^T,g=1,2,\ldots,G$ of length $N=N_F/G$, where $x_t^g(n) \in \left\lbrace 0,\mathcal{S} \right\rbrace,n=1,2,\ldots,N $. According to the corresponding $p_1=\lfloor \log_2\left(C\left(N,K\right)  \right)  \rfloor$ bits, only $K$ out of $N$ available subcarriers are selected as active by the index selector at each subblock $g$, while the remaining $N-K$ subcarriers are inactive and set to zero. On the other hand, the remaining $p_2=K \log_2(M)$ bits are mapped onto the considered $M$-ary signal constellation. Therefore, unlike classical MIMO-OFDM, $ \mathbf{x}_t,t=1,2,\ldots,T $ contains some zero terms whose positions carry information for MIMO-OFDM-IM.

% such as $ M $-ary phase-shift keying ($M$-PSK) or M-ary quadrature amplitude modulation ($M$-QAM) at each subblock.

In this study, active subcarrier index selection is performed by the reference look-up tables at OFDM index modulators of the transmitter. The considered reference look-up tables for $N=4,K=2$ and $N=4,K=3$ are given in Tables I and II, respectively, where $s_k\in \mathcal{S}$ for $ k=1,2,\ldots K $. As seen from Table I, for $N=4$ and $K=2$, the incoming $p_1=2$ bits can be used to select the indices of the two active subcarriers out of four available subcarriers according to the reference look-up table of size $C=2^{p_1}=4$. 

\begin{table}[!t]
	\begin{center}
		\setlength{\extrarowheight}{1pt}
		\caption{Reference Look-up Table for $N=4,K=2$ and $p_1=2$}
		%\vspace*{-0.3cm}
		%\setlength{\extrarowheight}{2pt}
		\label{tab:Look_up}
		\begin{tabular}[c]{|c||c||c|} \hline
			\textit{Bits} & \textit{Indices} & \textit{OFDM-IM subblocks $(\mathbf{x}_t^g)^T$}  \\ \hline \hline
			$[0\,\,\, 0]$ & $\left\lbrace 1, 3\right\rbrace $ & $\begin{bmatrix}s_{1} & 0 & s_2 & 0   \end{bmatrix}$  \\ \hline
			$[0\,\,\, 1]$ & $\left\lbrace 2,4\right\rbrace$ & $\begin{bmatrix}0 & s_1 & 0 & s_2   \end{bmatrix}$ \\ \hline
			$[1\,\,\, 0]$ & $\left\lbrace 1, 4\right\rbrace$ & $\begin{bmatrix} s_1 & 0 & 0 & s_2     \end{bmatrix}$ \\ \hline
			$[1\,\,\, 1]$ & $\left\lbrace 2,3\right\rbrace$ & $\begin{bmatrix} 0 & s_1 & s_2 &  0  \end{bmatrix}$  \\ \hline
		\end{tabular}
	\end{center}
	%\vspace*{-0.6cm}
\end{table}

\begin{table}[!t]
	\begin{center}
		\setlength{\extrarowheight}{1pt}
		\caption{Reference Look-up Table for $N=4,K=3$ and $p_1=2$}
		%\vspace*{-0.3cm}
		%\setlength{\extrarowheight}{2pt}
		\label{tab:Look_up}
		\begin{tabular}[c]{|c||c||c|} \hline
			\textit{Bits} & \textit{Indices} & \textit{OFDM-IM subblocks $(\mathbf{x}_t^g)^T$}  \\ \hline \hline
			$[0\,\,\, 0]$ & $\left\lbrace 1,2, 3\right\rbrace $ & $\begin{bmatrix}s_{1} & s_2 & s_3 & 0   \end{bmatrix}$  \\ \hline
			$[0\,\,\, 1]$ & $\left\lbrace 1,2,4\right\rbrace$ & $\begin{bmatrix}s_1 & s_2 & 0 & s_3   \end{bmatrix}$ \\ \hline
			$[1\,\,\, 0]$ & $\left\lbrace 1,3, 4\right\rbrace$ & $\begin{bmatrix} s_1 & 0 & s_2 & s_3     \end{bmatrix}$ \\ \hline
			$[1\,\,\, 1]$ & $\left\lbrace 2,3, 4 \right\rbrace$ & $\begin{bmatrix} 0 & s_1 & s_2 &  s_3  \end{bmatrix}$  \\ \hline
		\end{tabular}
	\end{center}
	%\vspace*{-0.7cm}
\end{table}

The OFDM index modulators in each branch of the transmitter obtain the OFDM-IM subblocks first and then concatenate these $G$ subblocks to form the main OFDM blocks $\mathbf{x}_t,t=1,2,\ldots,T$. In order to transmit the elements of the subblocks from uncorrelated channels, $G\times N$ block interleavers $(\Pi)$ are employed at the transmitter. The block interleaved OFDM-IM frames $\mathbf{\tilde{x}}_t,t=1,2,\ldots,T$ are processed by the inverse FFT (IFFT) operators to obtain 
$\mathbf{\tilde{q}}_t,t=1,2,\ldots,T$. We assume that the time-domain OFDM symbols are normalized to have unit energy, i.e., $ E\left\lbrace \mathbf{\tilde{q} }_t^H\mathbf{\tilde{q} }_t \right\rbrace=N_F  $ for all $ t $. After the addition of cyclic prefix of $C_p$ samples, parallel-to-serial and digital-to-analog conversions, the resulting signals sent simultaneously from $ T $ transmit antennas over a frequency selective Rayleigh fading MIMO channel, where $\mathbf{g}_{r,t}\in \mathbb{C}^{L\times 1}$ represents the $L$-tap wireless channel between the transmit antenna $ t $ and the receive antenna $ r $, whose elements are independent and identically distributed with $\mathcal{CN} (0,\frac{1}{L})$. Assuming the wireless channels remain constant during the transmission of a MIMO-OFDM-IM frame and $C_p > L$, after removal of the cyclic prefix and performing FFT operations in each branch of the receiver, the input-output relationship of the MIMO-OFDM-IM scheme in the frequency domain is obtained as
\begin{equation}
\mathbf{\tilde{y}}_r=\sum\nolimits_{t=1}^{T}\mbox{diag} \left( \mathbf{\tilde{x}}_t\right)  \mathbf{h }_{r,t} + \mathbf{w}_r 
\end{equation}
for $r=1,2,\ldots,R$, where $\mathbf{\tilde{y}}_r= \begin{bmatrix}
\tilde{y}_r(1) & \tilde{y}_r(2) & \cdots & \tilde{y}_r(N_F)
\end{bmatrix}^T$ is the vector of the received signals for receive antenna $r$, $\mathbf{h}_{r,t} \in \mathbb{C}^{N_F \times 1}$ represents the frequency response of the wireless channel between the transmit antenna $t$ and receive antenna $r$, and $\mathbf{w}_{r} \in \mathbb{C}^{N_F \times 1}$ is the vector of noise samples. The elements of $ \mathbf{h}_{r,t} $ and $\mathbf{w}_{r}$ follow $\mathcal{CN}\left(0,1 \right) $ and $\mathcal{CN}\left(0,N_{0,F} \right) $ distributions, respectively, where $N_{0,F}$ denotes the variance of the noise samples in the frequency domain, which is related to the variance of the noise samples in the time domain as $N_{0,T}=(N_F/(KG)) N_{0,F}$. We define the signal-to-noise ratio (SNR) as $\mathrm{SNR}=E_b/N_{0,T}$ where $E_b=(N_F+C_p)/m$ [joules/bit] is the average transmitted energy per bit. The spectral efficiency of the MIMO-OFDM-IM scheme is $mT/(N_F+C_p)$ [bits/s/Hz], which is equal to $T$ times that of the OFDM-IM scheme.

\section{Detection of MIMO-OFDM-IM Scheme}

After block deinterleaving in each branch of the receiver, the received signals are obtained for receive antenna $r$ as
\begin{equation}
\mathbf{y}_r=\sum\nolimits_{t=1}^{T}\mbox{diag} \left( \mathbf{x}_t\right)  \mathbf{\breve{h }}_{r,t} + \mathbf{\breve{w}}_r 
\label{eq:1}
\end{equation}
where $\mathbf{\breve{h }}_{r,t}$ and $\mathbf{\breve{w}}_r  $ are deinterleaved versions of $ \mathbf{h }_{r,t} $ and $\mathbf{w}_r  $, respectively. The detection of the MIMO-OFDM-IM scheme can be performed by the separation of the received signals in (\ref{eq:1}) for each subblock $g=1,2,\ldots,G$ as $ \mathbf{y}_r \!=\! \begin{bmatrix}
(\mathbf{y}_r^1)^T  \!& \cdots \!& (\mathbf{y}_r^G)^T 
\end{bmatrix}^T$, $ \mathbf{x}_t = \begin{bmatrix}
(\mathbf{x}_t^1)^T  \!& \cdots \!& (\mathbf{x}_t^G)^T 
\end{bmatrix}^T $, $\mathbf{\breve{h}}_{r,t} \!\! = \!\! \begin{bmatrix}
(\mathbf{\breve{h}}_{r,t}^1)^T  \!& \cdots \!& (\mathbf{\breve{h}}_{r,t}^G)^T \end{bmatrix}^T $, $\mathbf{\breve{w}}_{r} \! =\! \begin{bmatrix}
(\mathbf{\breve{w}}_{r}^1)^T \! & \cdots \! & (\mathbf{\breve{w}}_{r}^G)^T \end{bmatrix}^T $
%\begin{align}
%	\mathbf{y}_r &= \begin{bmatrix}
%		(\mathbf{y}_r^1)^T & (\mathbf{y}_r^2)^T & \cdots & (\mathbf{y}_r^G)^T 
%	\end{bmatrix}^T \nonumber\\
%	\mathbf{x}_t &= \begin{bmatrix}
%		(\mathbf{x}_t^1)^T & (\mathbf{x}_t^2)^T & \cdots & (\mathbf{x}_t^G)^T 
%	\end{bmatrix}^T \nonumber\\
%	\mathbf{\breve{h}}_{r,t} &= \begin{bmatrix}
%		(\mathbf{\breve{h}}_{r,t}^1)^T & (\mathbf{\breve{h}}_{r,t}^2)^ T & \cdots & (\mathbf{\breve{h}}_{r,t}^G)^T \end{bmatrix}^T \nonumber\\
%	\mathbf{\breve{w}}_{r} &= \begin{bmatrix}
%		(\mathbf{\breve{w}}_{r}^1)^T & (\mathbf{\breve{w}}_{r}^2)^ T & \cdots & (\mathbf{\breve{w}}_{r}^G)^T
%	\end{bmatrix}^T,
%\end{align}
for which we obtain
\begin{align}
\label{eq:3}
\mathbf{y}_r^g & =\sum\nolimits_{t=1}^{T}\mbox{diag} \left( \mathbf{x}_t^g\right)  \mathbf{\breve{h }}_{r,t}^g + \mathbf{\breve{w}}_r^g 
\end{align}
for $r=1,2,\ldots,R$, where $ \mathbf{y}_r^g= \begin{bmatrix}
	y_r^g(1) & y_r^g(2) & \cdots & y_r^g(N)
\end{bmatrix}^T$ is the vector of the received signals at receive antenna $r$ for subbblock $ g $, $\mathbf{x}_t^g=\begin{bmatrix}
x_t^g(1) & x_t^g(2) &\cdots &x_t^g(N)
\end{bmatrix}^T$ is the OFDM-IM subblock $g$ for transmit antenna $t$, $ \mathbf{\breve{h }}_{r,t}^g = \big[ \begin{matrix}
\breve{h}_{r,t}^g(1) & \breve{h}_{r,t}^g(2) &\cdots & \breve{h}_{r,t}^g(N)
\end{matrix} \big] ^T  $ and $ \mathbf{\breve{w}}_{r}^g = \begin{bmatrix}
\breve{w}_{r}^g(1) & \breve{w}_{r}^g(2) &\cdots & \breve{w}_{r}^g(N)
\end{bmatrix}^T  $. The use of the block interleaving ensures that $E  \big\lbrace \mathbf{\breve{h }}_{r,t}^g (\mathbf{\breve{h }}_{r,t}^g)^H  \big\rbrace = \mathbf{I}_{N} $, i.e., the subcarriers in a subblock are affected from uncorrelated wireless fading channels for practical values of $N_F$. 

A straightforward but costly solution to the detection problem of (\ref{eq:3}) is the use of maximum likelihood (ML) detector which can be realized for each subblock $g$ as 
\begin{equation}
\left( \mathbf{\hat{x}}_1^g,\ldots,\mathbf{\hat{x}}_T^g\right)= \arg\min_{\left( \mathbf{x}_1^g,...,\mathbf{x}_T^g\right) } \sum_{r=1}^{R}  \bigg\| \mathbf{y}_r^g -\sum_{t=1}^{T}\mbox{diag} \left( \mathbf{x}_t^g\right)  \mathbf{\breve{h }}_{r,t}^g\bigg\|^2.
\label{eq:ML} 
\end{equation} 
As seen from (\ref{eq:ML}), the ML detector has to make a joint search over all transmit antennas due the interference between the subblocks of different transmit antennas. Since $\mathbf{x}_t^g$ has $CM^K$ different realizations, the total decoding complexity of the ML detector in (\ref{eq:ML}), in terms of complex multiplications (CMs), is $\sim \mathcal{O}\left( M^{KT}\right) $ per subblock, which becomes impractical for higher order modulations and MIMO systems. Instead of the exponentially increasing decoding complexity of the ML detector, we propose a novel MMSE detection and LLR calculation based detector, which has a linear decoding complexity as that of classical MIMO-OFDM with MMSE detection.   

For the detection of the corresponding OFDM-IM subblocks of different transmit antennas, the following MIMO signal model is obtained from (\ref{eq:3}) for subcarrier $n$ of subblock $g$:
\begin{align}
\!\begin{bmatrix}
y_1^g(n) \\ y_2^g(n) \\ \vdots \\ y_R^g(n) \\ 
\end{bmatrix} \!\!\!&=\!\! \!
\begin{bmatrix}
\breve{h}_{1,1}^g(n) & \breve{h}_{1,2}^g(n) & \cdots & \breve{h}_{1,T}^g(n) \\
\breve{h}_{2,1}^g(n) & \breve{h}_{2,2}^g(n) & \cdots & \breve{h}_{2,T}^g(n) \\
\vdots & \vdots & \ddots & \vdots \\
\breve{h}_{R,1}^g(n) & \breve{h}_{R,2}^g(n) & \cdots & \breve{h}_{R,T}^g(n)
\end{bmatrix} \!\!\!\!
\begin{bmatrix}
x_1^g(n) \\ x_2^g(n) \\ \vdots \\ x_T^g(n) \\ 
\end{bmatrix} \!\!\!+\!\!\! \begin{bmatrix}
\!\breve{w}_1^g(n)\! \\ \!\breve{w}_2^g(n)\! \\ \!\vdots\! \\ \!\breve{w}_R^g(n)\! \\ 
\end{bmatrix} \nonumber \\
&\hspace*{2cm}\mathbf{\bar{y}}_n^g=\mathbf{H}_n^g \mathbf{\bar{x}}_n^g + \mathbf{\bar{w}}_n^g
\label{eq:MMSE}
\end{align}
for $n=1,2,\ldots,N$ and $g=1,2,\ldots,G$, where $ \mathbf{\bar{y}}_n^g $ is the received signal vector, $ \mathbf{H}_n^g $ is the corresponding channel matrix which contains the channel coefficients between transmit and receive antennas and assumed to be perfectly known at the receiver, $ \mathbf{\bar{x}}_n^g  $ is the data vector which contains the simultaneously transmitted symbols from all transmit antennas and can have zero terms due to index selection in each branch of the transmitter and $\mathbf{\bar{w}}_n^g$ is the noise vector. For classical MIMO-OFDM, the data symbols can be simply recovered after processing the received signal vector in (\ref{eq:MMSE}) with the MMSE detector. On the other hand, due to the index information carried by the subblocks of the proposed scheme, it is not possible to detect the transmitted symbols by only processing $ \mathbf{\bar{y}}_n^g $ for a given subcarrier $n$ in the MIMO-OFDM-IM scheme. Therefore, $N$ successive MMSE detections are performed for the proposed scheme using the MMSE filtering matrix \cite{MMSE}
\begin{equation}
\mathbf{W}_{n}^{g} = \left( \left( \mathbf{H}_n^g\right)^H  \mathbf{H}_n^g + \dfrac{\mathbf{I}_T}{\rho}  \right)^{-1} \left( \mathbf{H}_n^g\right)^H
\end{equation} 
\vspace*{-0.23cm}\\
for $n=1,2,\ldots,N$, where $\rho=\sigma_x^2 / N_{0,F}$, $\sigma_x^2=K/N$ and $E\big\lbrace \mathbf{\bar{x}}_n^g \left( \mathbf{\bar{x}}_n^g\right)^H \big\rbrace = \sigma_x^2 \mathbf{I}_T$ due to zero terms in $\mathbf{\bar{x}}_n^g$ come from the index selection. By the left multiplication of $ \mathbf{\bar{y}}_n^g $ given in (\ref{eq:MMSE}) with $ \mathbf{W}_n^g $, MMSE detection is performed as
\begin{equation}
\mathbf{z}_n^g=\mathbf{W}_n^g \mathbf{\bar{y}}_n^g=\mathbf{W}_n^g \mathbf{H}_n^g \mathbf{\bar{x}}_n^g + \mathbf{W}_n^g \mathbf{\bar{w}}_n^g,
\end{equation}    
where $\mathbf{z}_n^g=\begin{bmatrix}
z_n^g(1) & z_n^g(2) & \cdots & z_n^g(T)
\end{bmatrix}^T$ is the MMSE estimate of $ \mathbf{\bar{x}}_n^g $. The MMSE estimate of MIMO-OFDM-IM subblocks $\mathbf{\hat{x}}_t^g=\begin{bmatrix}
\hat{x}_t^g(1) & \hat{x}_t^g(2)  & \cdots & \hat{x}_t^g(N) 
\end{bmatrix}^T$ can be obtained by rearranging the elements of $\mathbf{z}_n^g,n=1,2,\ldots,N$ as $\mathbf{\hat{x}}_t^g=\begin{bmatrix}
z_1^g(t) & z_2^g(t)  & \cdots & z_N^g(t) 
\end{bmatrix}^T$
%\begin{equation}
%\mathbf{\hat{x}}_t^g=\begin{bmatrix}
%z_1^g(t) & z_2^g(t)  & \cdots & z_N^g(t) 
%\end{bmatrix}^T
%\end{equation}
for $t=1,2,\ldots,T$ and $g=1,2,\ldots,G$. As mentioned earlier, unlike classical MIMO-OFDM, $ \mathbf{\hat{x}}_t^g $ contains some zero terms, whose positions carry information; therefore, independent detection of the data symbols in $ \mathbf{\hat{x}}_t^g $ (with linear decoding complexity) is not a straightforward problem for the proposed scheme. As an example, rounding off individually the elements of $ \mathbf{\hat{x}}_t^g $ to the closest constellation points (the elements of $\left\lbrace 0,\mathcal{S} \right\rbrace $ for the proposed scheme) as in classical MIMO-OFDM may result a catastrophic active index combination that is not included in the reference look-up table, which makes the recovery of index selecting $p_1$ bits impossible.

In order to determine the active subcarriers in $ \mathbf{\hat{x}}_t^g $, the LLR detector of the proposed scheme calculates the following ratio which provides information on the active status of the corresponding subcarrier index $n$ of transmit antenna $t$:
\begin{equation}
\lambda_t^g(n)=\ln\dfrac{\sum_{m=1}^{M} P \left( x_t^g(n)=s_{m} \left. \right| \hat{x}_t^g(n) \right) }{P \left( x_t^g(n)=0 \left. \right| \hat{x}_t^g(n) \right)}
\label{eq:10}
\end{equation}
for $n=1,2,\ldots,N$, where $s_m \in \mathcal{S}$. Using Bayes formula and dropping the constant terms in (\ref{eq:10}), we obtain
\begin{equation}
\lambda_t^g(n)=\ln\dfrac{\sum_{m=1}^{M} P \left( \hat{x}_t^g(n) \left. \right|   x_t^g(n)=s_{m} \right) }{P \left( \hat{x}_t^g(n)  \left. \right|   x_t^g(n)=0 \right)}
\label{eq:LLR}
\end{equation}
which requires the conditional statistics of $\hat{x}_t^g(n)$ $(z_n^g(t))$. However, due to successive MMSE detection, the elements of $ \mathbf{\hat{x}}_t^g $ are still Gaussian distributed but have different mean and variance values. Let us consider the mean vector and covariance matrix of $\mathbf{z}_n^g$ conditioned on $x_t^g(n)\in\left\lbrace 0,\mathcal{S}\right\rbrace $, which are given as
\begin{align}
E\left\lbrace \mathbf{z}_n^g \right\rbrace &= \mathbf{W}_n^g \mathbf{H}_n^g E\left\lbrace  \mathbf{\bar{x}}_n^g  \right\rbrace = \left( \mathbf{W}_n^g \mathbf{H}_n^g\right)_t x_t^g(n) \nonumber \\
\mbox{cov}(\mathbf{z}_n^g) &=  \mathbf{W}_n^g \mathbf{H}_n^g \mbox{cov}\left(\mathbf{\bar{x}}_n^g  \right) \left(\mathbf{H}_n^g \right)^H  \!\! \left(\mathbf{W}_n^g \right)^H \!\!+\! N_{0,F}\mathbf{W}_n^g \left(\mathbf{W}_n^g \right)^H  
\label{eq:cov}
\end{align}
where $ E\left\lbrace  \mathbf{\bar{x}}_n^g  \right\rbrace $ is an all-zero vector except its $t$th element is $x_t^g(n)$, and
\begin{align}
 \mbox{cov}\left(\mathbf{\bar{x}}_n^g  \right) &= E\left\lbrace \left(\mathbf{\bar{x}}_n^g - E\left\lbrace \mathbf{\bar{x}}_n^g \right\rbrace  \right) \left(\mathbf{\bar{x}}_n^g - E\left\lbrace \mathbf{\bar{x}}_n^g \right\rbrace  \right)^H   \right\rbrace \nonumber \\
 &= \mbox{diag}\left( \begin{bmatrix}
 	\sigma_x^2 &  \ldots & \sigma_x^2 & 0 & \sigma_x^2 & \ldots & \sigma_x^2  
 \end{bmatrix}  \right) 	
\end{align}
is a diagonal matrix whose $t$th diagonal element is zero. From (\ref{eq:cov}), the conditional mean and variance of $\hat{x}_t^g(n)$ are obtained as 
\begin{align}
E\left\lbrace \hat{x}_t^g(n) \right\rbrace = \left( \mathbf{W}_n^g \mathbf{H}_n^g\right)_{t,t} x_t^g(n)  \nonumber \\
\mbox{var} \left(\hat{x}_t^g(n)  \right)  = \left(\mbox{cov}(\mathbf{z}_n^g) \right)_{t,t}. 
\label{eq:mean_variance}
\end{align}
Combining (\ref{eq:LLR}) and (\ref{eq:mean_variance}), the LLR for the $n$th subcarrier of $t$th transmitter for subblock g can be calculated as
\begin{equation}
\lambda_t^g(n)\!=\!\ln \!\left( \! \sum_{m=1}^{M} \! \exp\left(  - \frac{\left|\hat{x}_t^g(n)  \!- \!\left( \mathbf{W}_n^g \mathbf{H}_n^g\right)_{t,t} s_{m}\right|^2 }{\left(\mbox{cov}(\mathbf{z}_n^g) \right)_{t,t} }\right)  \right)\! + \frac{\left|\hat{x}_t^g(n) \right|^2 }{\left(\mbox{cov}(\mathbf{z}_n^g) \right)_{t,t}}
\label{eq:15}
\end{equation}
for $n=1,2,\ldots,N$, $t=1,2,\ldots,T$ and $g=1,2,\ldots,G$. After the calculation of $N$ LLR values for a given subblock $g$ and transmit antenna $t$, in order to determine the indices of the active subcarriers, the LLR detector calculates the following LLR sums for $c=1,2,\ldots,C$ according to the look-up table: $d_t^g(c)= \sum\nolimits_{k=1}^{K} \lambda_t^g(i^c_k)$, 
%\begin{equation}
%d_t^g(c)= \sum\nolimits_{k=1}^{K} \lambda_t^g(i^c_k)
%\end{equation}
where $\mathcal{I}^c=\left\lbrace i^c_{1},i^c_{2},\ldots,i^c_{K} \right\rbrace $ denotes the possible active subcarrier index combinations. As an example, for Table I, we have $\mathcal{I}^1=\left\lbrace 1,3 \right\rbrace$, $ \mathcal{I}^2=\left\lbrace 2,4 \right\rbrace  $, $ \mathcal{I}^3=\left\lbrace 1,4 \right\rbrace  $ and $ \mathcal{I}^4=\left\lbrace 2,3 \right\rbrace  $. The LLR detector determines the active subcarriers for a given subblock $g$ and transmit antenna $t$ as $\hat{c}=\arg\max_c d_t^g(c)$ and  $\mathcal{\hat{I}}_t^g = \left\lbrace i^{\hat{c}}_{1}, i^{\hat{c}}_{2},\ldots,i^{\hat{c}}_{K} \right\rbrace$. The $M$-ary symbols transmitted by the active subcarriers are determined with ML detection as
\begin{equation}
\hat{s}_t^g(k)=\arg\min_{s_m \in \mathcal{S}}	\Big|\hat{x}_t^g(i^{\hat{c}}_k)-\left( \mathbf{W}^g_{i^{\hat{c}}_k}\mathbf{H}^g_{i^{\hat{c}}_k}\right)_{t,t} s_m \Big|^2 
\label{eq:17}
\end{equation} 
for $k=1,2,\ldots,K$, where the metrics in (\ref{eq:17}) were calculated in (\ref{eq:15}) and do not increase the decoding complexity. After this point, index selecting $p_1$ bits are recovered from the look-up table and $M$-ary symbols are demodulated to obtain the corresponding $p_2$ information bits. The total number of CMs performed in (6)-(\ref{eq:17}) for the MIMO-OFDM-IM scheme is $2T^3+5T^2R+T(R+M+1)$ per subcarrier while this value is equal to $T^3+2T^2R+T(R+M)$ for classical MIMO-OFDM, where the decoding complexity of both schemes increases linearly with respect to $M$ $(\sim\mathcal{O}(M))$ due to MMSE detection.

%\vspace*{-0.1cm}
\section{Simulation Results}
In this section, we provide computer simulation results for MIMO-OFDM-IM and classical V-BLAST type MIMO-OFDM schemes employing BPSK, QPSK and $16$-QAM $(M=2,4\mbox{ and }16)$  modulations and MMSE detection. We consider three different $T\times R$ MIMO configurations: $2\times 2,4\times 4$ and $8\times 8$. The following OFDM parameters are assumed in all Monte Carlo simulations: $N_F=512,C_p=16,L=10$. 

In Fig. 2, we compare the BER performance of the proposed MIMO-OFDM-IM scheme for $N=4,K=2$ with classical MIMO-OFDM for $M=2$ at same spectral efficiency values. As seen from Fig. 2, the proposed scheme provides significant BER performance improvement compared to classical MIMO-OFDM, which increases with higher order MIMO systems. As an example, the MIMO-OFDM-IM scheme achieves approximately $10$ dB better BER performance than classical MIMO-OFDM at a BER value of $10^{-5}$ for the $8\times 8$ MIMO system.

In Figs. 2 and 3, we extend our simulations to higher spectral efficiency values and compare the BER performance of the proposed MIMO-OFDM-IM scheme $(N=4,K=3)$ with classical MIMO-OFDM for $M=4$ and $16$, respectively. As seen from Figs. 2 and 3, the proposed scheme still maintains its advantage over classical MIMO-OFDM in all considered configurations. It is interesting to note that the proposed scheme has the potential to achieve close or better BER performance than the reference scheme, even using a lower order MIMO system in most cases.

%\vspace*{-0.1cm}
\section{Conclusions and Future Work}
A novel scheme called MIMO-OFDM with index modulation has been proposed as an alternative multicarrier transmission technique for 5G networks. It has been shown via extensive computer simulations that the proposed scheme can provide significant BER performance improvements over classical MIMO-OFDM for several different configurations. The following points remain unsolved in this study: i) performance analysis, ii) the selection of optimal $N$ and $K$ values, iii) diversity techniques for MIMO-OFDM-IM, and iv) implementation scenarios for high mobility.

\bibliographystyle{IEEEtran}
\bibliography{IEEEabrv,MIMO_OFDM_IM_R1}

\vspace*{3cm}
\begin{figure}[!h]
	\begin{center}\resizebox*{12cm}{10cm}{\includegraphics{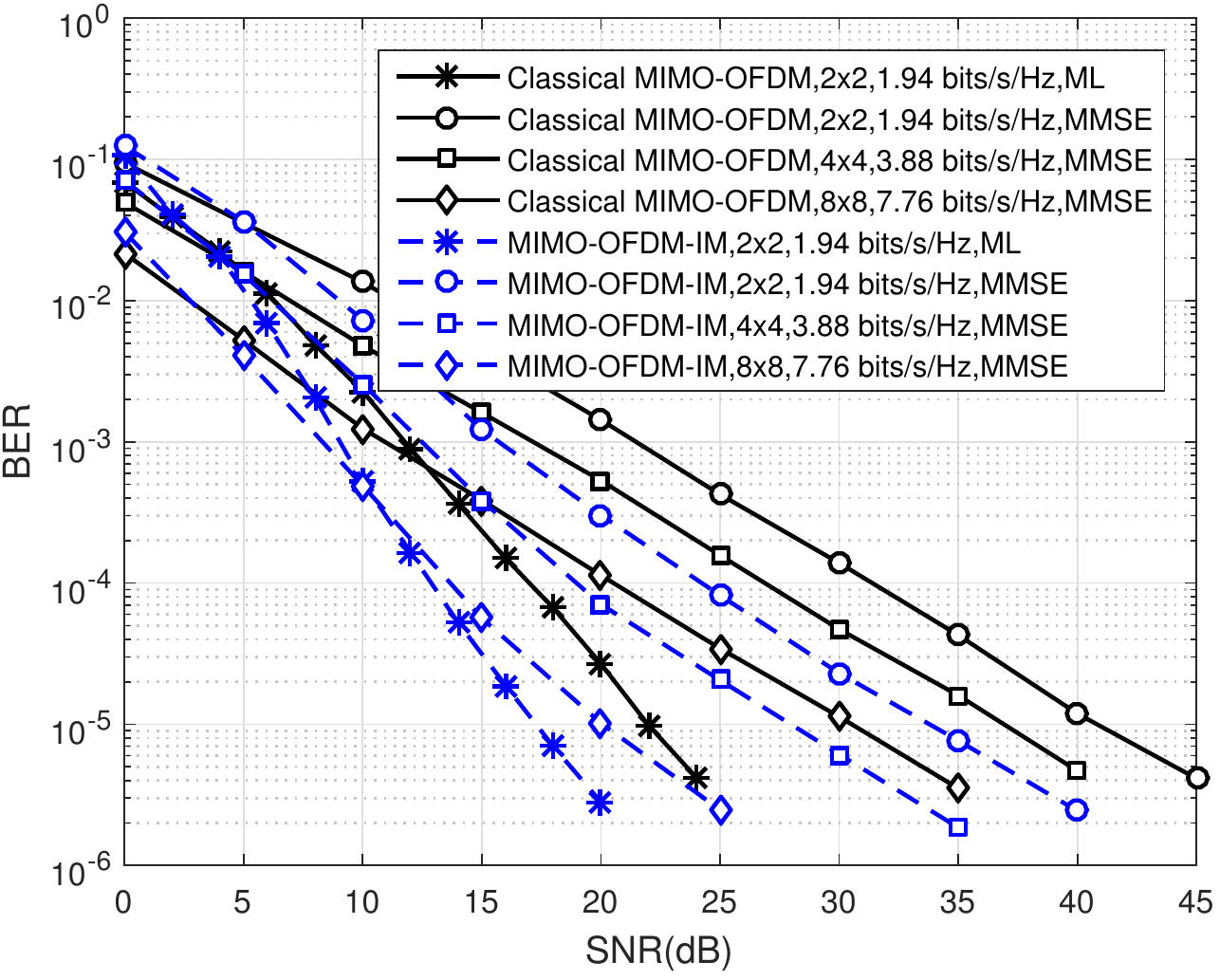}}
		\caption{Performance comparison of MIMO-OFDM and MIMO-OFDM-IM $(N=4,K=2)$ for BPSK modulation $(M=2)$, MMSE/ML detection}
	\end{center}
\end{figure}

\begin{figure}[!t]
	\begin{center}\resizebox*{12cm}{10cm}{\includegraphics{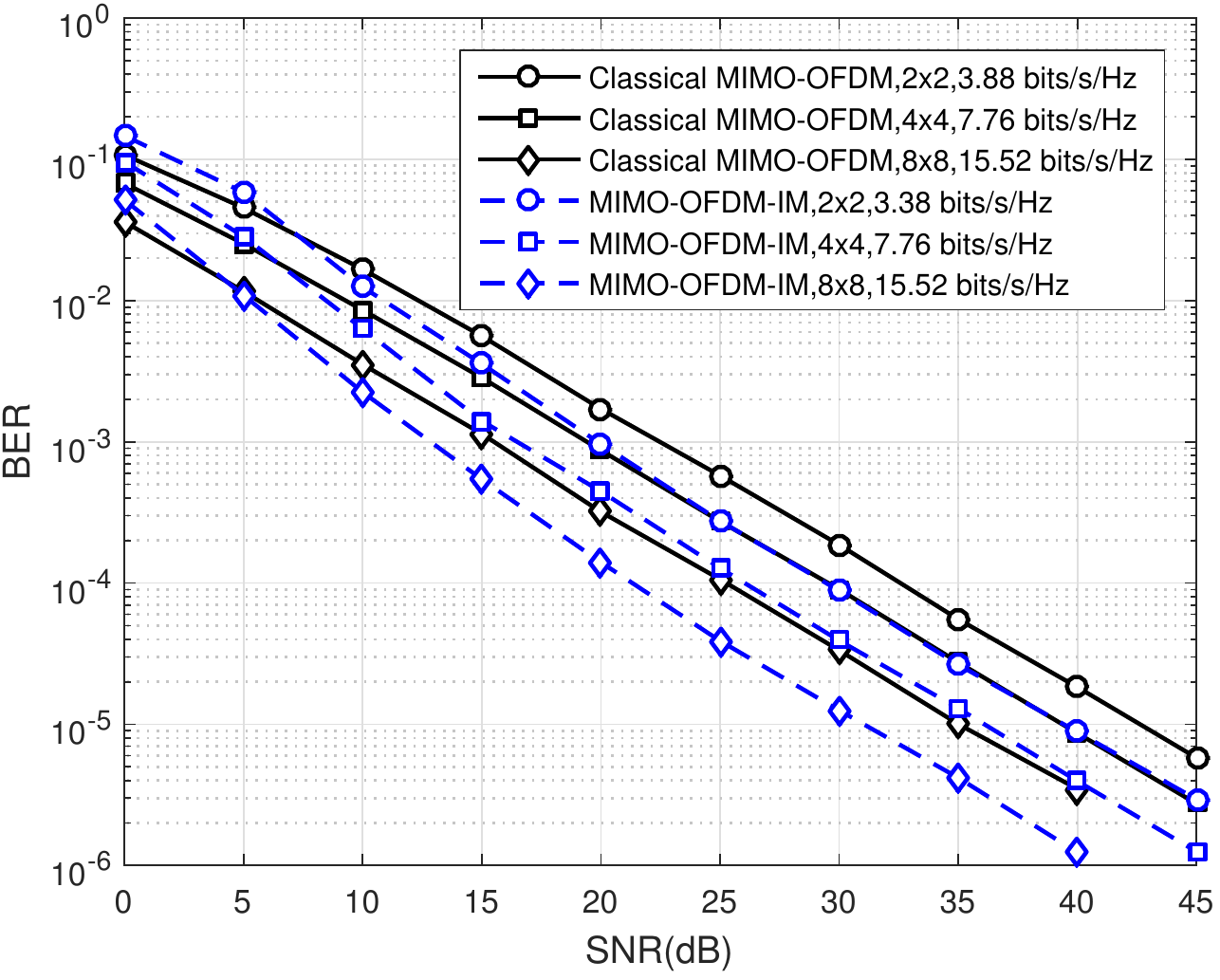}}
		\caption{Performance comparison of MIMO-OFDM and MIMO-OFDM-IM $(N=4,K=3)$ for QPSK modulation $(M=4)$, MMSE detection}
	\end{center}
\end{figure}

\begin{figure}[!t]
	\begin{center}\resizebox*{12cm}{10cm}{\includegraphics{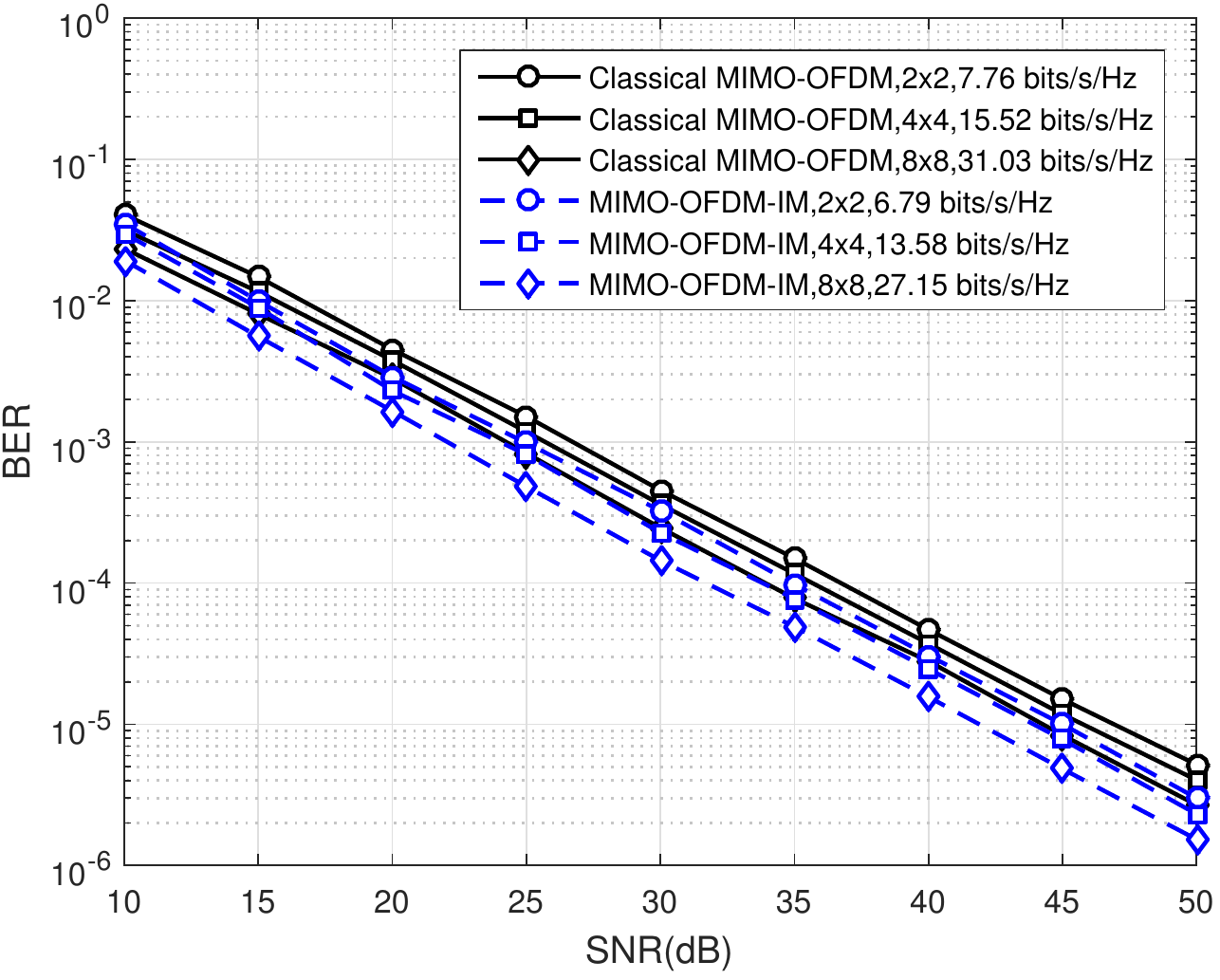}}
		\caption{Performance comparison of MIMO-OFDM and MIMO-OFDM-IM $(N=4,K=3)$ for $16$-QAM $(M=16)$, MMSE detection}
	\end{center}
\end{figure}

%\newpage

% that's all folks
\end{document}